\documentclass[pra, 12 pt]{revtex4}
\usepackage[english]{babel}
\usepackage{amsmath}
\usepackage{epsfig}

\begin{document}


\title{ELEMENTARY EXCITATIONS AND HEAT CAPACITY SINGULARITY IN SUPERFLUID HELIUM}

\author{V.B.Bobrov$^2$, S.A.Trigger$^{1,2}$}
\address{$^1$Joint\, Institute\, for\, High\, Temperatures, Russian\, Academy\,
of\, Sciences, 13/19, Izhorskaia Str., Moscow\, 125412, Russia;\\
$^2$ Eindhoven  University of Technology, P.O. Box 513, MB 5600
Eindhoven, The Netherlands;
emails:\, vic5907@mail.ru,\;satron@mail.ru}

\begin{abstract}
We proceed from the premise that the spectrum of elementary excitations in the normal component in Landau's theory of superfluidity should depend on the superfluid helium temperature. This leads to generalization of the Landau superfluidity criterion. On this basis, taking into account available experimental data on inelastic neutron scattering, it is shown that, in addition to phonon--roton excitations, there is one more type of elementary excitations in superfluid helium, which we called helons. The energy spectrum with such a momentum dependence was first proposed by Landau. The helon energy spectrum shape and its temperature dependence make it possible to explain the singular behavior of the heat capacity of superfluid helium near its phase transition to the normal state.

PACS number(s): 67.25.de, 47.37.+q, 67.25.dj, 67.10.Fj\\

\end{abstract}

\maketitle

According to the seminal phenomenological Landau theory [1,2], superfluid helium is a liquid consisting of superfluid and normal components. The superfluid component moves without friction and is not involved in the energy transport in the form of heat. The normal component moves with friction and is involved in heat transport. In this case, according to the Landau theory [1], the normal component is a gas of elementary excitations which are characterized by the dependence of the energy spectrum $\varepsilon(p)$ on the momentum $p$. If the flow velocity $V$ of the superfluid component reaches the critical velocity $V_{cr}$, determined from the condition
\begin{eqnarray}
V_{cr}=min \left(\varepsilon(p)/p\right), \label{A1}
\end{eqnarray}
superfluidity breakdown occurs. Thus, the superfluidity phenomenon cannot to be observed at velocities $V>V_{cr}$. This statement [1] known as the Landau superfluidity criterion is in qualitative agreement with experimental data on superfluid helium motion in capillaries (see [3] for more details). To quantitatively describe the superfluid helium motion in capillaries, the superfluid component inhomogeneity caused by boundary effects (see, e.g., [4]) should be taken into account. In this paper, we consider the infinite medium model corresponding to the thermodynamic limit.

Thus, the Landau superfluidity criterion is in fact the superfluidity breakdown criterion, since it initially assumes the existence of superfluidity. Otherwise, considering that there are well defined acoustic elementary excitations (phonons) in any liquid, we would obtain a nonzero critical velocity equal to the speed of sound in a corresponding liquid. To clarify the problem it should be taken into account that the elementary excitation spectrum $\varepsilon$ in the normal component is a function of not only the momentum $p$, but also thermodynamic parameters of the system under consideration, e.g., temperature $T$, i.e.,
\begin{eqnarray}
\varepsilon=\varepsilon(p;T).\label{A2}
\end{eqnarray}
Hence, the critical velocity $V_{cr}$ determined from relation (1) is also a function of thermodynamic parameters, $V_{cr}=V_{cr}(T)$. Let us further take into account that the superfluidity phenomenon is absent at the temperature $T>T_\lambda$, where $T_\lambda$ is the superfluid transition temperature, i.e., liquid is normal. Therefore, it should be accepted that
\begin{eqnarray}
V_{cr}(T>T_\lambda)=0 .\label{A3}
\end{eqnarray}
Thus, we can formulate the generalized Landau superfluidity criterion exactly
as the superfluidity criterion, rather than the superfluidity breakdown criterion, in the following form: if the spectrum of elementary excitations in liquid satisfies the conditions
\begin{eqnarray}
V_{cr}(T)>0 \qquad \mbox{for} \qquad T<T_\lambda; \qquad V_{cr}(T)=0 \qquad \mbox{for}\qquad T>T_\lambda, \label{A4}
\end{eqnarray}
then the corresponding liquid at temperatures $T<T_\lambda$ is superfluid; the superfluidity breakdown occurs at velocities $V>V_{cr}$.

As noted above, there are well defined acoustic elementary excitations in any liquid, both normal and superfluid one; therefore, the phonon spectrum of elementary excitations $\varepsilon(p)=cp$ where $c$ is the speed of sound, does not satisfy the generalized Landau superfluidity criterion (4). This means that one more branch of elementary excitations should exist in addition to phonons, which differs essentially from the phonon spectrum. Thus, we should introduce one more correction to the formulation of the generalized Landau superfluidity criterion, associated with the fact that several "branches"\, of elementary excitations can exist in liquid. Therefore, among all possible values of the critical velocity $V^\alpha_{cr}(T)$ determined by each spectrum (spectrum index $\alpha$), our interest is in only that providing a minimum value among $V^\alpha_{cr}(T)$. Hence, the quantity $V_{cr}(T)$, appearing in relation (4), is determined from the condition
\begin{eqnarray}
V_{cr}(T)=min_{\alpha}V^\alpha_{cr}(T); \qquad V^\alpha_{cr}(T)= min \left(\varepsilon^\alpha(p,T)/p\right).\label{A5}
\end{eqnarray}
We note that, according to (4) and (5), two cases are possible:\\
- either there is an excitation branch with $V^\alpha_{cr}=0$ in normal liquid, which differs essentially from the phonon spectrum and, during the transition to the superfluid state, yields $V^\alpha_{cr}>0$,\\
- or there is one more branch of elementary excitations in the normal component of superfluid liquid, which differs essentially from the phonon spectrum and disappears at temperatures $T>T_\lambda$.

Before turning to the discussion of the possible shape of the energy spectrum of the additional branch of elementary excitations differing essentially from phonons, let us consider the situation with the phonon--roton spectrum of elementary excitations $\varepsilon^{ph-rot}(p)$ (see Fig. 1) in superfluid helium, which was proposed by Landau in [2]. The shape of the phonon--roton spectrum of elementary excitations was confirmed in experiments on inelastic neutron scattering in superfluid helium (see, e.g., [5,6].

Furthermore, numerous experiments on inelastic neutron scattering (see, e.g., [7]-[9]) show that the phonon--roton spectrum of elementary excitations very weakly depends on temperature to $T_\lambda =2.17$ K for all values of momenta, including phonon and roton spectral regions. Moreover, phonon--roton excitations also exist at temperatures $T>T_\lambda$, where liquid helium is in the normal state [10]. Thereby, according to the above discussion, there is reason to believe that the phonon--roton spectrum of elementary excitations is not nearly related to the explanation of the superfluidity phenomenon in liquid helium. This point of view is confirmed by the experimental results on inelastic neutron scattering in liquid metals (Fig. 1), where the phonon--roton spectrum of elementary excitations was detected (see, e.g., [11]-[14], which was noticed in [15]. Similar excitations were also experimentally detected in the two-dimensional Fermi liquid [16].
\begin{figure}[h]
\centering\includegraphics[width=6cm]{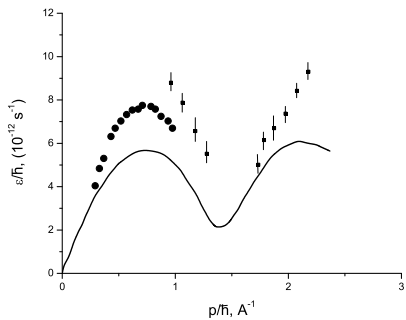}
\caption{{\protect\footnotesize {The position of the maximum in the dynamic ion structure factor in liquid rubidium at 320 K: full circles - the experimental data of Copley and Rowe [11], vertical bars - the experimental data of Glaeser et al. [12], solid curve - the numerical calculation of the dispersion equation [13].}}}\label{Fig.1}
\end{figure}


Let us now pay attention that Landau in his first paper [1] proposed to consider, in addition to phonons, elementary excitations which he initially called "rotons"\ with the spectrum
\begin{eqnarray}
\varepsilon^r=\Delta^r+\frac{p^2}{2\mu^r}.\label{A6}
\end{eqnarray}
Let us refer this type of excitations to as "Landau rotons"\, in contrast to later introduced rotons [2]  being a fraction of the single phonon--roton branch of excitations.
In (6), $\Delta^r$ is the energy of the Landau roton with the effective mass $\mu^r$ at the zero momentum $p=0$.
It is clear that the critical velocity $V^r_{cr}$ of Landau\ rotons, according to (1), (5), and (6) is given by
\begin{eqnarray}
V_{cr}^r=\sqrt{2\mu^r \Delta^r}.\label{A7}
\end{eqnarray}
Assuming that $\Delta^r$ in (6) depends on temperature and satisfies the condition
\begin{eqnarray}
\Delta^r (T)=0 \qquad \mbox{for}\qquad T>T_\lambda, \label{A8}
\end{eqnarray}
the spectrum of such excitations satisfies the generalized Landau superfluidity criterion (4).

Let us pay attention that, while satisfying condition (8), elementary excitations with energy spectrum (6) exist only in superfluid helium, in contrast to the phonon--roton spectrum characteristic of liquid. To distinguish elementary excitations with the spectrum (6)--(8) from "rotons" (used in the literature) in the phonon--roton spectrum and vortical "Landau rotons"\ currently dropped from consideration, and taking into account that the vortical nature of these excitations in the infinite medium is not obvious, in what follows, we refer these elementary excitations to as "helons" (index $h$).

The existence of helons with the spectrum (6)--(8) is in fact confirmed by experiments on inelastic neutron scattering [17,18] (see Fig. 2), in which, in addition to the maxima in the dynamic structure factor of superfluid helium, corresponding to the phonon--roton spectrum of elementary excitations, the maxima were detected, whose positions appeared close to the spectrum of the free helium atom $\varepsilon^a(p)=p^2/2m $ (here $m$ is the helium atom mass) for the region of the momentum transferred values $q>0.5 {\AA}^{-1}$ . The corresponding experimental data were called the spectrum of "single-atom scattering". It is clear that the spectrum of the free helium atom $\varepsilon^{(a)}(p)$ (as well as other spectra with the same $p$-dependence at small $p$) does not satisfy the generalized Landau superfluidity criterion (4), (5). Therefore, the new experiments are required for the smaller values of $q=p/\hbar$ and for different temperatures, which can prove the existence of helons and give estimation of the value $\Delta^{(h)}(T)$.

\begin{figure}[h]
\centering\includegraphics[width=6cm]{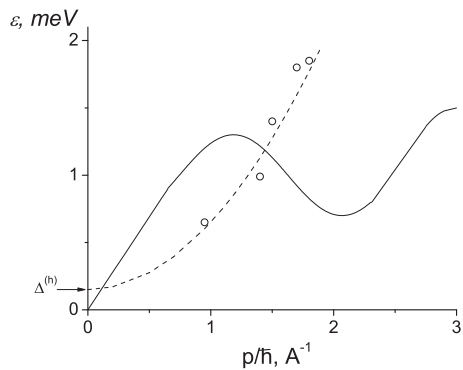}
\caption{{\protect\footnotesize {The position of dynamical structure factor maximum of superfluid helium at 1.2 K. The continuous line is the experimental phonon-roton spectrum [17], the circles are the positions of "quasi-free" maxima from [17],  the dotted line shows the assumed curve for the helon excitations in HeII.}}}\label{Fig.2}
\end{figure}


In attempts to theoretically explain the experimentally observed maxima in the dynamic structure factor of superfluid helium, corresponding to the "single-atom scattering" spectrum, the possible existence of helons was not taken into consideration (see, e.g.,[20]-[22] and references therein). We also note that theoretical models were repeatedly proposed in microscopic descriptions of superfluid helium, in which the spectrum of elementary excitations, similar to the spectrum of helons arises (see, e.g.,[23]-[26] and references therein); however, the existence of the corresponding maximum in the dynamic structure factor has not yet been confirmed in these models.

To provide condition (8), there is an appropriate quantity in the Landau theory, i.e., the superfluid component density $n_s$ for which the condition
\begin{eqnarray}
n_s (T>T_\lambda)=0 \qquad \mbox{at}\qquad T>T_\lambda\label{A9}
\end{eqnarray}
is satisfied. Then we can assume that $\Delta^{(h)} \simeq [n_s]^\gamma$,\,$\gamma>0$ at $T<T_\lambda$.

In this case, for dimensionality reasons, to determine the quantity $\Delta^{(h)}(T)$, several quantities with energy dimension can be constructed, based on the superfluid component density $n_s$, in particular, $\hbar^2 n_s^{2/3}(T)/m$ and $\hbar^2 L n_s(T)/m$, where $L$ is the so-called scattering length which is completely defined by the interparticle interaction potential of helium atoms.

Thus, there is good reason to believe that, in addition to phonon--roton elementary excitations, there are helons with spectrum (6)--(8) in the normal component of superfluid helium in the absence of boundary effects. Let us consider consequences from this statement. According to the Landau superfluidity theory [1,2], the free energy per unit volume of superfluid helium at temperature $T$ can be written as
\begin{eqnarray}
F=E_0+\sum_a F^{(a)}, \qquad F^{(a)}= T \int \frac{d^3 p}{(2\pi\hbar)^3}\ln\left\{ 1-\exp\left[-\frac{\varepsilon^{(a)}(p;T)}{T}\right]\right\}.   \label{A10}
\end{eqnarray}
Here $E_0$ is the ground state energy per unit volume of superfluid helium, which depends only on its density $n$ equal to the sum of the densities of superfluid and normal components, $n=n_N+n_s$. The quantity $F^{(a)}$ is the free energy per unit volume of superfluid helium, corresponding to elementary excitations of type $a$ with energy spectra $\varepsilon^{(a)}(p;T)$ and corresponding to helons (6)--(8) and phonon--roton excitations.

It immediately follows from (10) that the average (internal) energy per unit volume of superfluid helium is given by
\begin{eqnarray}
F=E_0+\sum_a E^{(a)}, \qquad E^{(a)}= T \int \frac{d^3 p}{(2\pi\hbar)^3} \frac{\varepsilon^{(a)}(p;T)-T[\partial \varepsilon^{(a)}(p;T)/\partial T]}{\exp\left[\varepsilon^{(a)}(p;T)/T\right]-1}.   \label{A11}
\end{eqnarray}
In turn, from (11), it is easy to verify that the heat capacity $c_V = (\partial E/\partial T)_V$ of superfluid helium, by virtue of condition (8), has a peculiarity at the temperature $T=T_\lambda$ for the helon energy spectrum, caused by the temperature dependence of $\Delta^{(h)}(T)$:
\begin{eqnarray}
\lim_{T\rightarrow (T_\lambda-0)}c_V=\infty, \qquad \mbox{при} \qquad  \lim_{T\rightarrow (T_\lambda-0)}\Delta^{(h)} (T)= 0, \qquad \lim_{T\rightarrow (T_\lambda-0)}\frac{d \Delta^{(h)} (T)}{d T} \neq 0 \label{A12}
\end{eqnarray}
which is widely known in the literature as the $\lambda$-curve of the heat capacity.

The simple calculation for the singular part of the heat capacity
below $T_\lambda$ leads to the temperature dependence
\begin{eqnarray}
c_V\rightarrow\frac {A}{\sqrt{1-\left(\frac{T}{T_\lambda}\right)^\gamma}}\label{A13}
\end{eqnarray}
where $A= \gamma^2 [2\mu^{(h)} \Delta^{(h)}(T=0)]^{3/2}/8\pi\hbar^3$. Here we assumed the form of temperature dependence of the value $\Delta^{(h)}(T)=\Delta^{(h)}(T=0) [1-(T/T_\lambda)^\gamma]$ with $\gamma>0$ and used the expansion  of the expression under the integral for $c_V$ for small values of $p$, suggested in [27]. The value $\Delta^{(h)}(T=0)$ is the functional of the interaction potential between the particles and tends to zero when interaction disappears. In this case singularity is absent and the heat capacity is continuous with the well known fracture at $T=T_\lambda$.  Probably, in the transition point there is an infinite jump of the heat capacity. These assumption based on disappearance of helons at $T=T_\lambda$, as well as on the character of the experimental data in vicinity of the transition point (see e.g. [3], [28], [29]). Explanation of the behavior of the heat capacity above $T_\lambda$ requires microscopical consideration of the correlation effects in quantum liquid.

Let us pay attention that the phonon--roton spectrum of excitations does not exhibit a similar anomaly of the specific heat $c_V$ even taking into account its temperature dependence due to the linearity of the phonon--roton spectrum at small momenta, $\varepsilon^{ph-rot}(p\rightarrow 0)\rightarrow c p$.

Thus, according to the above consideration, in addition to elementary excitations with the phonon--roton energy spectrum in superfluid helium, there are helons (6)--(8) which satisfy the generalized Landau superfluidity criterion. The consideration of the temperature dependence of the helon energy spectrum allows explanation of the anomalous behavior of the specific heat of superfluid helium in the vicinity of the phase transition to the normal state.

\section*{Acknowledgment}
This study was supported by the Netherlands Organization for Scientific Research (NWO) and the Russian Foundation for Basic Research, projects no. 12-08-00822-a and no. 12-02-90433-Ukr-a.\\

\end{document}